%% file: ms.tex
\documentclass{an}

\usepackage{graphicx}
\usepackage{aas_macros}
\usepackage{times}
\usepackage{amsmath}
\include{definitions}

\begin{document}

%
%
\Pagespan{228}{}
\Yearpublication{2007}%
\Yearsubmission{2006}%
\Month{3}%
\Volume{328}%
\Issue{3-4}%
\DOI{10.1002/asna.200610724}%

\title{Linear Sensitivity of Helioseismic Travel Times to Local Flows}
\author{A.C. Birch \inst{1}\fnmsep\thanks{Corresponding author:
  \email{aaronb@cora.nwra.com}\newline}
\and L. Gizon\inst{2}
}
\titlerunning{Local Flows}
\authorrunning{A.C. Birch \& L. Gizon}
\institute{
NWRA, CoRA Division, 3380 Mitchell Lane, Boulder CO 80301 USA
\and 
Max-Planck-Institut f\"{u}r Sonnensystemforschung, Max-Planck-Stra{\ss}e 2, 37191 Katlenburg-Lindau, Germany
}
\received{30 May 2005}
\accepted{11 Nov 2005}
\publonline{later}


\abstract{Time-distance helioseismology is a technique for measuring the time for waves to travel from one point on the solar surface to another.  These wave travel times are affected by advection by subsurface flows.  Inferences of plasma flows based on observed travel times depend critically on the ability to accurately model the effects of subsurface flows on time-distance measurements.  We present a Born-approximation based computation of the sensitivity of time-distance travel times to weak, steady, inhomogeneous subsurface flows. Three sensitivity functions are obtained, one for each component of the 3D vector flow.
We show that the depth sensitivity of travel times to horizontally uniform flows is given approximately by the kinetic energy density of the oscillation modes which contribute to the travel times.  For flows with strong depth dependence, the Born approximation can give substantially different results than the ray approximation. }

\maketitle

\section{Introduction}

\sloppy Time-distance helioseismology (\cite{Duvall1993}) is a  technique for measuring the time for waves to travel from one point on the solar surface to another.  Subsurface flows advect waves and as a result alter the observed travel times.  Thus wave travel times can be used as probes of subsurface flows  (e.g.\ \cite{Kosovichev1997,Zhao2004})

One of the key steps in the interpretation of travel times is to estimate the effect of subsurface flows on travel times.  The ray approximation (\cite{Kosovichev1997}), in which travel-time shifts are only caused by inhomogeneities located along the ray path connecting the observation points,  has been employed in many time-distance studies of plasma flows (e.g.\ \cite{Kosovichev1997, Zhao2001, Zhao2003,Zhao2004}).  An alternative to the ray approximation is the first Born approximation (e.g.\ \cite{Birch2000}; \cite{Gizon2002}, in the context of time-distance helioseimology).  The first Born approximation takes into account a single scattering and thus flows located away from the ray path can affect the travel time.

\cite{Birch2004b} studied the ranges of validity of the Born and ray approximations in a toy problem consisting of jets in a homogeneous two-dimensional medium.  This study showed that there are flow configurations for which the Born approximation is valid while the ray approximation is not, especially when the transverse size of the jet is much smaller than the wavelength. As a result, we would like to use the first Born approximation to compute the sensitivity of travel times to flows for use in time-distance helioseismology.

Here we employ the Born approximation approach of Gizon \& Birch (2002), referred to as GB02 hereafter, to compute the sensitivity of travel times to weak, steady, three-dimensional subsurface flows in the Sun.  We use the phenomenological model of Birch et al.\ (2004), referred to as B04 hereafter, to describe how waves are excited and damped by convection.  We work in Cartesian geometry, which is appropriate for waves that travel distances much less than the solar radius and also have wavelengths much smaller than the solar radius.

The remainder of this paper is organized as follows. In \S\ref{sec.kernels} we derive the general expression for the sensitivity of travel times to weak flows. In \S\ref{sec.examples} we show a few example calculations.  We compare the Born and ray approximations in \S\ref{sec.compare}.  We conclude in \S\ref{sec.discuss} with a summary and a short discussion of the implications of the work presented here.

\section{Sensitivity Functions}\label{sec.kernels}
In this section we use the Born approximation to obtain the linear sensitivity of travel times to weak and steady subsurface flows.  We are looking for kernels ${\bf K}=(K_x,K_y,K_z)$ which satisfy
\begin{equation}
\delta\tau(\bx_1,\bx_2) = \intr_\odot\id\br\; {\bf K}(\br;\bx_1,\bx_2)\cdot\vel(\br) \; ,
\end{equation}
where the integration variable $\br$ runs over the entire volume of the solar model and $\vel=(v_x,v_y,v_z)$ is the flow field. Throughout this paper we will denote three-dimensional position vectors by $\br=(\bx,z)$ where $\bx=(x,y)$ is the horizontal position vector and $z$ is depth.  The travel-time difference between surface locations $\bx_1$ and $\bx_2$ is denoted as  $\delta\tau(\bx_1,\bx_2)$ and defined by
\begin{equation}
\delta\tau(\bx_1,\bx_2) = \tau_+(\bx_1,\bx_2) - \tau_+(\bx_2,\bx_1) \; ,
\end{equation}
where $\tau_+(\bx_i,\bx_j)$ is the one-way travel time, as defined by GB02, from $\bx_i$ to $\bx_j$.

Following GB02 and B04,  we begin by considering damped and driven solar oscillations with a displacement field $\bxi$ that obeys, to lowest order in the flow velocity $\vel$,
\begin{equation}
\label{eq.wave_equation}
\left[\bL_0+\delta\bL\right]\bxi = \bS \; , 
\end{equation}
with the wave equation operator in the absence of flows, $\bL_0$ (e.g.\ \cite{Bell1967}), given by
\begin{eqnarray}
\label{eq.define_L}
\bL_0\bxi &=& \rho_0 \ddot{\bxi} - \bnabla\left[\gamma p_0 \bnabla\cdot\bxi+\bxi\cdot\bnabla p_0\right]  \\
&& +(\bnabla\cdot\bxi)\bnabla p_0 + \bxi\cdot\bnabla(\bnabla p_0) + \rho_0\partial_t\left(\Gamma\bxi\right)\, , \nonumber
\end{eqnarray}
where $\rho_0$, $p_0$, and $\gamma$ are the background density, pressure, and ratio of specific heats. We use solar model S (\cite{jcd96}). The damping operator is $\Gamma$ (B04). In equation~(\ref{eq.define_L}) we have used the Cowling approximation and also neglected the variation of the gravitational acceleration with depth. We have also neglected the Coriolis force.  The source function ${\bf S}$ is intended to represent the driving of oscillations by near-surface turbulent convection.  In order to compute the covariance of the wavefield $\bxi$ we only need the covariance of the source ${\bf S}$. We choose to use the source covariance described by B04.

The first-order perturbation to the wave equation introduced by a flow (e.g.\ \cite{Bell1967}) is given by
\begin{equation}
\delta\bL\bxi = 2\rho_0\partial_t\vel\cdot\bnabla\bxi \, .
\end{equation}
This term only captures the direct advection of waves by the flow $\vel$.  Any associated changes in the background (non-wave) density and sound speed are neglected. The problem of the sensitivity of travel-times to changes in sound speed has already been addressed by B04.  Density perturbations could be treated in essentially the same manner.

The observations used for time-distance helioseismology are typically time series of images of the line-of-sight Doppler velocity near the solar surface.  For the sake of simplicity we assume that the line-of-sight is vertical.  In addition, we approximate the Doppler velocity as the time derivative of the displacement at a fixed geometrical height $z_{\rm obs}=200$~km.  In this case, the observed wavefield, $\phi(\bx,t)$, is related to the wave displacement by
\begin{equation}
\label{eq.define_phi}
\phi(\bx,t) = \filter \left\{ \partial_t \xi_z(\bx,t,z_{\rm obs}) \right\}\; ,
\end{equation} 
where the function $\filter$ denotes the action of the instrument point-spread function and any filters applied during the data analysis (e.g.\ phase-speed filters).  In the Fourier domain, we can write the wavefield as
\begin{equation}
\phi(\bk,\omega) = -i\omega F(\bk,\omega) \xi_z(\bk,\omega,z_{\rm obs}) \; ,
\end{equation}
where $\bk$ is the horizontal wavevector, $\omega$ the angular frequency, and $F(\bk,\omega)$ represents the filter in the Fourier domain.  Throughout this paper we will employ the Fourier convention of GB02, for a function $f$ of horizontal position and time, we have
\begin{equation}
f(\bx,t) = \int\id\bk\id\omega\; f(\bk,\omega) e^{i\sbk\cdot\sbx-i\omega t} \; .
\end{equation}
We will employ the same symbol for functions and their Fourier transforms, e.g.\ $f(\bk,\omega)$ denotes the transform of $f(\bx,t)$.




In order to compute travel-time kernels we need first to obtain Green's functions.  We define Green's functions $\bG^j$ as the solutions to
\begin{equation}
\label{eq.green_function}
\bL^0 \bG^j(\bx,t,z,z') = \hat{{\bf e}}_j\delta(\bx)\delta(t)\delta(z-z') \, ,
\end{equation}
where $\hat{{\bf e}}_j$ with $j=x,y,z$ are unit vectors along the coordinate axes. We use the same boundary conditions as in B04: zero Lagrangian pressure at the top of model S and no vertical motion at a depth of 300~Mm below the photosphere (this bottom boundary condition has essentially no effect on the computations presented here).  The Green's vector  $\bG^j(\bx,t,z,z')$ gives the displacement response as a function of horizontal position $\bx$, time $t$, and height $z$, to a delta function source in the $\hat{{\bf e}}_j$ direction at horizontal position ${\bf 0}$, time $t=0$, and height $z'$.  We use the normal-mode summation approximation solution to equation~(\ref{eq.green_function}) given by B04.

We now have all of the ingredients to compute travel-time kernels using the recipe of equations~(26) and~(32) from GB02.  The result for the kernels for velocity is
\begin{equation}
\!\!\! K_j(\br;\bx_1,\bx_2) = 4\pi {\rm Re} \!\! \int_0^\infty\!\!\!\!\id\omega\;\!\! W^*_{\rm diff}(\omega) {\cal C}_j(\br;\bx_2|\bx_1;\omega) \;,
\end{equation}
with $j=x,y,z$.  The function $W_{\rm diff}$ is the linear sensitivity of the travel-time to changes in the cross-covariance and is defined in  GB02.  The sensitivity of the cross-covariance to flows in the $j$ direction is denoted by ${\cal C}^j$ and given by
\begin{eqnarray}
{\cal C}_j(\br;\bx_2|\bx_1;\omega) && =  2 (2\pi)^7  i \omega^3 \rho_0(z) m(\omega) \\
&\times& \left[ {\bf\II}^j(\bx-\bx_1,z,\omega)\cdot{\bf\I}(\bx-\bx_2,z,\omega) \right. \nonumber \\
&+& \left. {\bf\II}^{j*}(\bx-\bx_2,z,\omega)\cdot{\bf\I}^*(\bx-\bx_1,z,\omega)\right] \nonumber \, .
\end{eqnarray}
where $\br=(\bx,z)$. The function $m(\omega)$ is the source autocorrelation function defined in B04. The vectors ${\bf I}(\bx)$ and ${\bf II}(\bx)$ we define in terms of their horizontal Fourier transforms:
\begin{eqnarray}
{\rm \I}^j(\bk,z,\omega) &=&  F(-\bk,\omega) G_z^{j}(-\bk,\omega,z_{\rm obs},z) , \\
{\bf\II}^j(\bk,z,\omega) &=&  F(\bk,\omega) H_z^*(\bk,\omega,z_{\rm obs}) \partial_j {\bf H}^*(\bk,\omega,z)  .
\end{eqnarray}
The Green's vector ${\bf H}$ is defined as
\begin{equation}
\label{eq.green_function_src}
{\bf H}(\bk,\omega,z) = \partial_{z'} {\bf G}^z (\bk,\omega,z,z')|_{z'=z_{\rm src}} \; ,
\end{equation}
where the source depth $z_{\rm src}$ is chosen to be 100~km below the photosphere (see B04 for a discussion of the source model). The general mathematical structure of the kernels was explained by GB02.

\section{Example Calculations}\label{sec.examples}

In this section we show the results of two example calculations.  The first example is for $p_1$ travel-time differences for the travel distance $\Delta=\| \bx_2 - \bx_1 \| = 7$~Mm obtained using a phase-speed filter.  The second example is for surface gravity wave travel-time differences at a travel distance of $\Delta=10$~Mm. 

\subsection{$p_1$ ridge}
\label{sec.examplesp}

For the first example, the filter function $F(k,\omega)$ is given as the product of three separate filters,
\begin{equation}
F(\bk,\omega) = F_1(k,\omega) F_2(k,\omega) {\rm OTF}(k) \; .
\end{equation}
The filter $F_1$ removes the $f$-mode ridge and frequencies below 1.5~mHz and above 5~mHz.  The phase-speed filter, $F_2$, is given by
\begin{equation}
F_2(\bk,\omega) = e^{-(w/k-v_{\rm p})^2 / 2\delta v^2_p } \; ,
\end{equation}
with $k=\|\bk\|$, $v_p=12.8$~km/s and $\delta v_p=2.6$~km/s (this is filter 1 from Couvidat et al. 2006). This filter isolates a section of the $p_1$ ridge.
Finally a filter ${\rm OTF}(\bk) = e^{-\alpha k}$ with $\alpha=1.75$~Mm is used as a very rough approximation for the optical transfer function of the MDI/{\it SOHO} high-resolution observing mode.

\begin{figure*}
\includegraphics[width=6in]{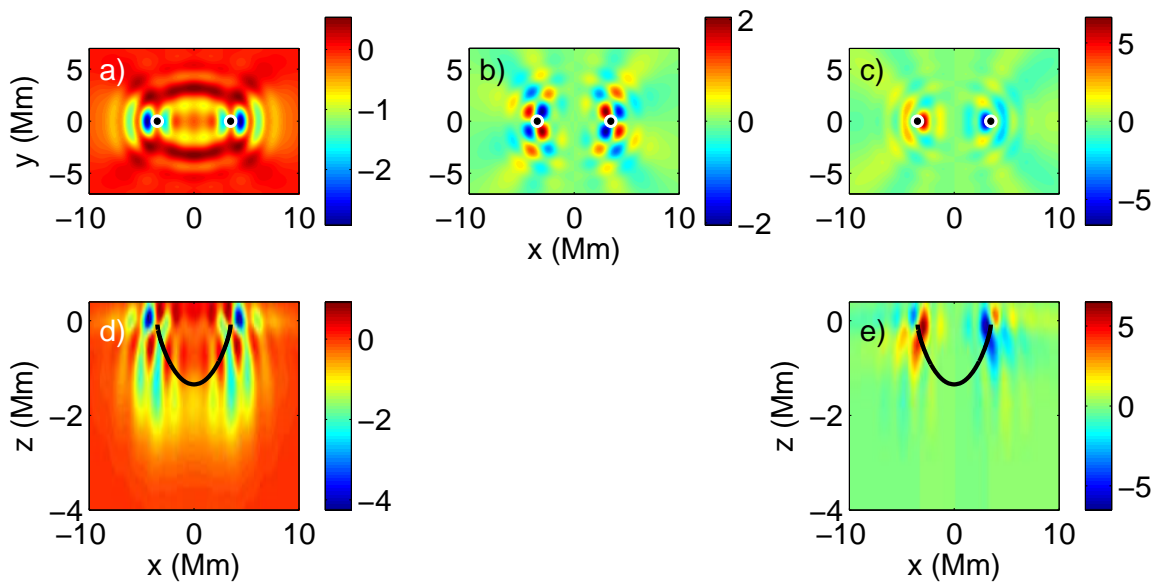}
\caption{Slices through an example of the sensitivity of a $p_1$ travel-time difference to local flows.  Panels~(a),~(b), and~(c) are horizontal slices, at the photosphere, through the kernels $K_x$, $K_y$, and $K_z$ respectively.  The symmetries of these three are different. The kernel $K_x$ is symmetric in both $x$ and $y$.  The kernel $K_y$ is symmetric in $x$ and anti-symmetric for $y$.
The kernel $K_z$ is anti-symmetric $x$ and symmetric in $y$.  Because of these symmetries, only $K_x$ has a non-zero total integral, as a result the travel-time difference $\delta\tau(\bx_1,\bx_2)$ is not sensitive (at first order) to uniform vertical flows or uniform flows in the cross-ray path direction. Panel~(d) shows a slice through $K_x$ at $y=0$.  The heavy black line shows the ray path.  Notice that the $p_1$ mode structure is seen in depth.  Panel~(e) shows a slice through $K_z$ at $y=0$, again with the ray path shown as the heavy black line.  By symmetry, the kernel $K_y$ is zero at $y=0$. In all panels the units are s~Mm$^{-3}$/(km/s). }
\label{fig.pmode_five_panel}
\end{figure*}

Figure~\ref{fig.pmode_five_panel} shows slices through the components of the kernel ${\bf K}$ for travel-time differences, for the $p_1$ case.  Figure~\ref{fig.pmode_five_panel}a shows a slice through $K_x$ at the photosphere. This kernel is symmetric in both $x$ and $y$. As with the kernels shown by GB02 and B04, ellipse- and hyperbola-shaped  features are visible.  The ringing in the horizontal directions is a result of the finite band-width of the wavefield.  Figures~\ref{fig.pmode_five_panel}b and~\ref{fig.pmode_five_panel}c show horizontal slices through the kernels $K_y$ and $K_z$ at the photosphere.  Different symmetries are visible in these slices; $K_y$ is anti-symmetric in both $x$ and $y$, while $K_z$ is symmetric in $y$ and antisymmetric in $x$.  Because of these symmetries both $K_y$ and $K_z$ integrate to zero.  The kernel $K_x$ has a non-zero total integral; spatially uniform flows in the $\hat{\bx}$ direction cause travel-time differences.

Figure~\ref{fig.pmode_five_panel}d shows a vertical slice through $K_x$ at $y=0$.  Also shown is the ray path corresponding to a frequency of 4.5~mHz.   The maximum height of the ray path is limited by the upper turning point (80~km below the photosphere).  For small travel distances $\Delta$, such as the example shown here, the lower turning point of the ray path is frequency dependent. In this case we have chosen to compute a ray at 4.5~mHz, which is the frequency where the wavefield has maximum power (after filtering). The $p_1$ mode structure is visible in $K_x$.  There is one maximum in sensitivity near the photosphere, and another near the lower turning point of the mode. This depth dependence will be discussed in later in this section.
     
A slice through $K_z$ at $y=0$ is shown in Figure~\ref{fig.pmode_five_panel}e.   The kernel $K_z$ is largest where the ray is mostly vertical, as expected from the ray approximation.

\subsection{Surface gravity waves}
\label{sec.examplesf}

For the second example we compute the sensitivity of $f$-mode travel-time differences to flow, for the case of a travel distance $\Delta=10$~Mm.
In this case the filter function $F$ was chosen to be
\begin{equation}
F(k,\omega) = F_3(k,\omega) {\rm OTF}(k)
\end{equation}
The filter $F_3$ selects only the $f$-mode and removes all $p$-modes and frequencies below 1.5~mHz and above 5~mHz.

\begin{figure*}
\includegraphics[width=6in]{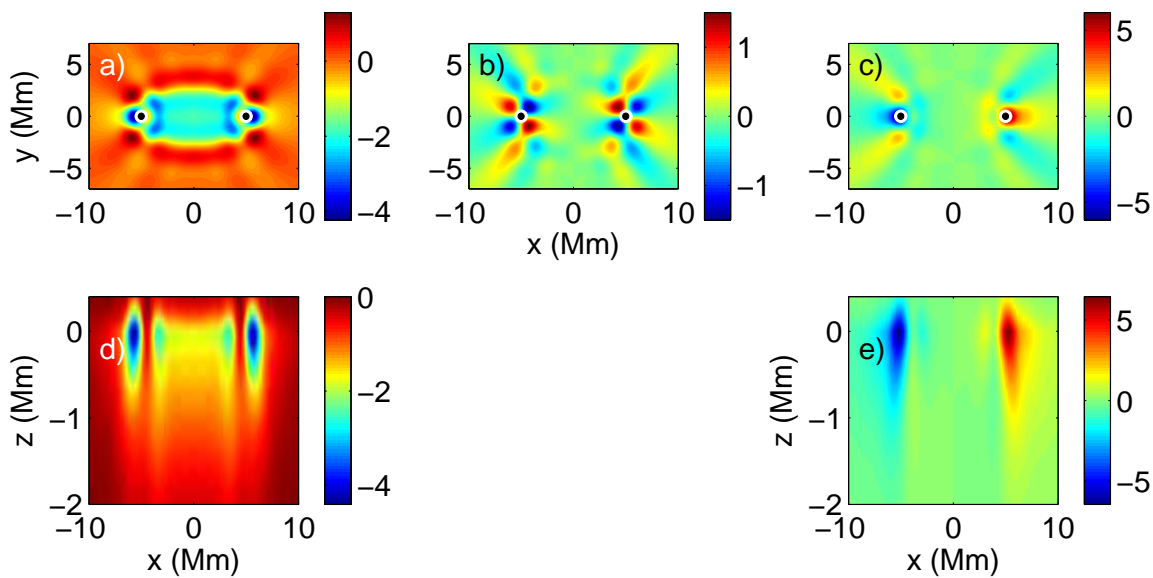}
\caption{Slices through the sensitivity of a $f$-mode travel time to local flows.  Panels~(a),~(b), and~(c) are horizontal slices, at the photosphere, through the kernels $K_x$, $K_y$, and $K_z$ respectively. 
Panel~(d) shows a slice through $K_x$ at $y=0$.  Panel~(e) shows a slice through $K_z$ at $y=0$.  By symmetry, the kernel $K_y$ is zero at $y=0$.    In all panels the units are s~Mm$^{-3}$/(km/s). } 
\label{fig.fmode_five_panel}
\end{figure*}

Figure~\ref{fig.fmode_five_panel} shows the results of the example $f$-mode calculation.
The kernels $K_x$, $K_y$, and $K_z$ show the same symmetries as in the $p$-mode case shown in Figure~\ref{fig.pmode_five_panel}.  As a result the kernel $K_x$ has a non-zero total integral, while $K_y$ and $K_z$ both integrate to zero.

A horizontal slice at the photosphere through $K_x$ is shown in Figure~\ref{fig.fmode_five_panel}a.
Notice that the $K_x$ kernel is the three-dimensional version of the kernel shown by Gizon et al. (2000) and in more detail by Jackiewicz et al.\ (2006).

\begin{figure}[t!]
\includegraphics[width=3in]{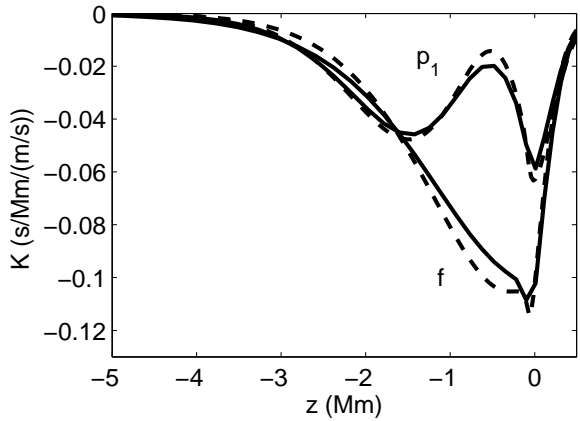}
\caption{Horizontal integrals of the $K_x$ kernels for the $p_1$ and $f$ cases (solid lines) 
shown in figures~\ref{fig.pmode_five_panel} and~\ref{fig.fmode_five_panel}.  Also shown are scaled kinetic energy densities (dashed lines) for the modes at the dominant wavenumber. }
\label{fig.depth}
\end{figure}

Figure~\ref{fig.depth} shows the depth dependence of the sensitivity functions shown in Figures~\ref{fig.pmode_five_panel} and~\ref{fig.fmode_five_panel}, for the case of horizontally uniform flows.  In both cases, the depth dependence is roughly proportional to the kinetic energy density of the associated mode.  This assumption was suggested for the $f$-mode by Gizon \& Duvall (2000b).


\section{Comparison with Ray Theory}\label{sec.compare}

As described in the introduction, the ray approximation (\cite{Kosovichev1997}) has been used to predict the travel-time shifts caused by sub-surface flows.  In this section we compare the predictions of the ray and Born approximations.  We will consider here very simple models of flows at supergranular scales.  We choose to study cylindrically symmetric flow fields, $\vel(r,z) = v_z(r,z) \hat{\bf z} + v_r(r,z)\hat{\bf r}$, of the form:
\begin{eqnarray}
\label{eq.flow1} v_r(r,z) &=& a f(r) h(z)  \; ,\\
\label{eq.flow2} v_z(r,z) &=& \frac{a}{r}\partial_r\left[r f(r)\right] n(z) \; ,
\end{eqnarray}
where $r$ is the distance from the symmetry axis of the flow and $z$ is depth.
The horizontal variation of the radial flow is given by
\begin{equation}
f(r) = J_0(kr) e^{-r/L} \; ,
\end{equation}
where $L$ is the decay length of the flow away from the center of the cell, and $k$ is the wavenumber associated with the radial variation in the flow.  The depth variation of the radial velocity we choose as
\begin{equation}
h(z) = e^{- (z-z_{\rm t})^2/D_1^2} - \beta e^{ - (z-z_{\rm b})^2 / D^2_2 } ,
\end{equation}
where $z_{\rm t}$ and $z_{\rm b}$ are the depths at the top and bottom of the cell, $D_1$ is the vertical scale of the outflow component of the flow, $D_2$ is the vertical scale of the inflow component, and the coefficient $\beta$ is chosen so that no vertical flows through the cell are required by mass conservation.  The depth dependence, $n(z)$, of the vertical flow we then choose so that the flow satisfies mass conservation
\begin{equation}
\partial_z \left[ \rho_0(z) n(z) \right] =\rho_0(z) h(z) \; , 
\end{equation}
with the upper boundary condition $n(z)=0$ at $z=z_{\rm t}$ where $z_t$ is the top of the cellular flow.

\begin{figure}
\includegraphics[width=\linewidth]{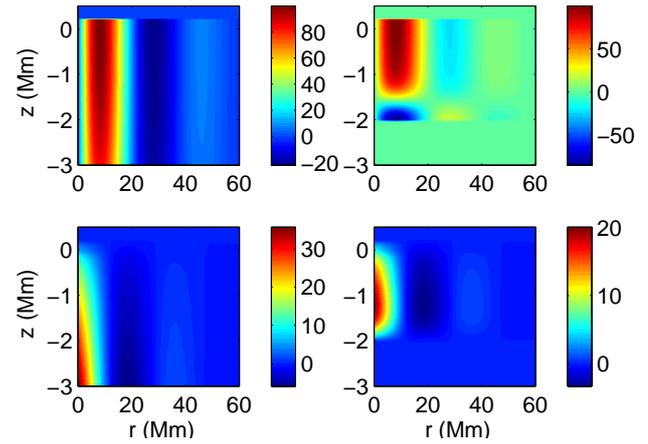}
\caption{Radial (top panels) and vertical (bottom panels) flows for model A (left column) and model B (right column).
The units of the color scales are  m/s.  The axis of symmetry of the cellular flow is $r=0$ and the photosphere is at $z=0$. Upflows correspond to positive values of the vertical velocity.}
\label{fig.flows}
\end{figure}

We consider two models.  Both have $(L,k)=(20~{\rm Mm},0.18~{\rm Mm}^{-1})$.  Model A has $(z_{\rm t}, z_{\rm b}, D_1, D_2) = (0.2,-8, 8,1)$~Mm and model B has $(z_{\rm t}, z_{\rm b}, D_1, D_2) = (0.2,-2,4,0.5)$~Mm.  In both cases we choose the amplitude $a$ so that the maximum radial flow speed is 100~m~s$^{-1}$.  Model A represents a deep flow that has very little vertical variation near the photosphere.  Model B represents a shallow flow.  The radial and vertical flows for these two models are shown in figure~\ref{fig.flows}.



For both models we compute the travel-time differences $\delta\tau(\bx_1,\bx_2)$ with $(\bx_1,\bx_2)=(x-\Delta/2,x+\Delta/2)\hat{\bx}$, where $\Delta$ is the travel distance and $\hat{\bx}$ is the unit vector in the $x$ direction.   For the ray-approximation travel times we use equation~(15) from Kosovichev \& Duvall (1997) with ray paths computed according to equation~(11) of that paper, for the travel distance $\Delta=7$~Mm (the same distance as for the kernels shown in Figure~\ref{fig.pmode_five_panel}).  We compute Born approximation travel-time shifts for the $p$-mode case described in \S\ref{sec.examplesp}.

\begin{figure}
\includegraphics[width=3in]{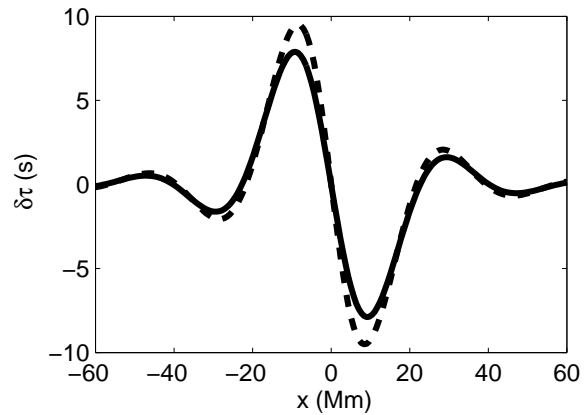}
\caption{Travel-time differences in the ray approximation (dashed line) and Born approximation (solid line) for model A ``deep flow''.   For this particular flow, the maximum difference between the two approximations is 2~s, about 25\%.}
\label{fig.ray_bornA}
\end{figure}

Figure~\ref{fig.ray_bornA} shows the ray and Born approximation travel-time differences for model A, the deep cellular flow.  The two approximations give travel-time shifts that are similar to within about two seconds.  In the Born approximation, the contribution of $\hat{\bz}$ component of the flow is about 5\% of the contributions from the $x$ component of the flow and the contribution from the $\hat{\by}$ component of the flow contributes less then one percent of the signal.    We note here that we have used, in the ray approximation, only a single ray at 4.5~mHz (this is the frequency where the wavefield has maximum power).  It may be that by taking a weighted average of many ray travel times, the ray approximation could be brought into agreement with the Born approximation.  A study of this procedure is beyond the scope of this simple example.

\begin{figure}
\includegraphics[width=3in]{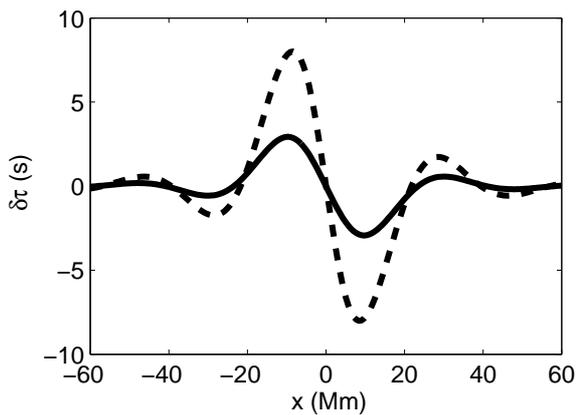}
\caption{Comparison between the ray approximation (dashed line) and Born approximation (solid line) for travel-time differences caused by the model B ``shallow flow''. For this flow, the Born approximation and the ray approximation give travel-time shifts that differ by up to five seconds. }
\label{fig.ray_bornB}
\end{figure}

Figure~\ref{fig.ray_bornB} shows the ray and Born approximation travel-time differences for model B, the shallow flow.  For this case, the ray and Born approximations give substantially different results.  The main cause of this difference is that the ray approximation is not sensitive to flows that are below the lower turning point of the ray, while in the Born approximation the sensitivity extends below the lower turning point. In this particular example, the strong counter-flow is sensed in the Born approximation, but not in the ray approximation.

\section{Discussion}\label{sec.discuss}

We employed the Born approximation to obtain the three-dimensional sensitivity of time-distance measurements to advection by local flows.  In this paper we addressed the important question of time-independent mass flows.  We have shown that for horizontally uniform flows the depth dependence of the sensitivity of travel times is given approximately by the kinetic energy density of the mode which contributes most to the travel times.

For simple cylindrically symmetric models of supergranulation-scale convection cells we showed that the Born and ray approximations can give results that are substantially different when the flow varies in the depth range just below the lower turning point of the ray.  This suggests that for inversions of supergranulation-scale flows it may be important to use kernels based on the Born approximation rather than the ray approximation.

In the future, a number of improvements could be implemented: inclusion of  modes above the acoustic cutoff frequency, taking  spherical geometry into account, and treatment of time-dependent flows.


\acknowledgements

The work of ACB was supported by NASA contracts NNH04CC05C and NNH06CD84C. It is planned to make this kernel caculation code available through the HELAS Network.

\end{document}

%% file: definitions.tex
\newcommand{\by}{\mbox{{\boldmath$y$}}}
\newcommand{\bz}{\mbox{{\boldmath$z$}}}

\newcommand{\intr}{\int\!\!\!\!\int\!\!\!\!\int}

\newcommand{\I}{{I}} 
\newcommand{\II}{{I}\!{I}} 
\newcommand{\sbk}{\mbox{{\boldmath\scriptsize$k$}}} 
\newcommand{\sbx}{\mbox{{\boldmath\scriptsize$x$}}} 
	
\newcommand{\bk}{\mbox{{\boldmath$k$}}}	
\newcommand{\vel}{\mbox{{\boldmath$v$}}} 
\newcommand{\bxi}{\mbox{{\boldmath$\xi$}}} 
\newcommand{\bnabla}{\mbox{{\boldmath$\nabla$}}} 
\newcommand{\bS}{\mbox{{\boldmath$S$}}}	
\newcommand{\bL}{\mbox{{\boldmath$\mathcal{L}$}}} 

\newcommand{\bG}{\mbox{{\boldmath$G$}}} 
\newcommand{\id}{\mbox{$\rm d$}} 
\newcommand{\bx}{{\mbox{{\boldmath$x$}}}} 
\newcommand{\br}{\mbox{{\boldmath$r$}}} 
\newcommand{\filter}{\mathcal{F}} 

%% file: ms.bbl
\begin{thebibliography}{}
\bibitem[Birch et al. 2004]{Birch2004} Birch, A.~C., {Kosovichev}, A.~G. \& {Duvall}, T.~L, Jr. : 2004 ApJ~608, 580
\bibitem[Birch \& Felder (2004)]{Birch2004b} Birch, A.~C. \& Felder, G. : 2004, ApJ~616, 1261
\bibitem[Birch \& Kosovichev 2000]{Birch2000}  Birch, A.~C. \& Kosovichev, A.~G. : 2000 Sol. Phys.~192, 193
\bibitem[Christensen-Dalsgaard et al.\ 1996]{jcd96}  Christensen-Dalsgaard, J. \& others 1996: Science~272, 1286
\bibitem[Couvidat et al. 2006]{Couvidat2006} Couvidat, S., Birch, A.~C., Kosovichev, A.~.G.: 2006, ApJ~640, 516
\bibitem[Duvall et al. 1993]{Duvall1993} Duvall, T.~L, Jr, Jefferies, S.~M, Harvery, J.~W., \& Pomerantz, M.~A.: 1993, Nature~362, 430
\bibitem[Gizon \& Birch 2002]{Gizon2002}  Gizon, L. \& Birch, A.~C.: 2002, ApJ~571, 966
\bibitem[Gizon et al. 2000]{Gizon2000} Gizon, L., Duvall, T.L., Jr. \& Larsen, R.M. 2000: JApA~21, 339
\bibitem[Gizon \& Duvall 2000]{Gizon2000b} Gizon, L., Duvall, T.L., Jr. : 2000b, Sol. Phys.~192, 177
\bibitem[Jackiewicz et al.(2006)]{Jackiewicz2006}
Jackiewicz, J., Gizon, L., Birch, A.~C.:  2006, in Proc SOHO 18, ESA SP-624  
\bibitem[Kosovichev \& Duvall 1997]{Kosovichev1997} Kosovichev, A.~G. \& Duvall, Jr. , T.~L. 1997 : "Acoustic tomography of solar convective flows and structures", in: SCORe'96 : Solar Convection and Oscillations and their Relationship, 241
\bibitem[Lynden-Bell \& Ostriker 1967]{Bell1967} Lynden-Bell, D. \& Ostriker, J.~P. 1967 : MNRAS~136, 293
\bibitem[Zhao et al. 2001]{Zhao2001} Zhao, J., Kosovichev, A.G., Duvall, T.~L., Jr.: 2001, ApJ~557, 384
\bibitem[Zhao \& Kosovichev 2003]{Zhao2003} Zhao, J., Kosovichev, A.G.: 2003, ApJ~591, 446
\bibitem[Zhao et al. 2004]{Zhao2004} Zhao, J., Kosovichev, A.G., Duvall, T.~L., Jr.: 2004, ApJ~607, L135
\end{thebibliography}
